\documentstyle[aps,prl,preprint,tighten,eqsecnum]{revtex}
\begin{document}
\title
{Finite temperature properties of quantum antiferromagnets
in a uniform magnetic field in one and two dimensions}
\author{\large Subir Sachdev, T. Senthil and R. Shankar}
\address{Departments of Physics and Applied Physics, P.O. Box 208120,\\
Yale University, New Haven, CT 06520-8120}
\date{\today}
\maketitle
\begin{abstract}
Consider a $d$-dimensional
antiferromagnet with a
quantum disordered ground state and a gap to bosonic excitations with
non-zero spin. In a finite external magnetic field,
this antiferromagnet will undergo
a phase transition to a ground state with non-zero magnetization,
describable as the condensation of a dilute gas of bosons. The finite
temperature properties of the Bose gas in the vicinity of this
transition are argued to obey a
hypothesis of {\em zero scale-factor universality\/} for $d <2$,
with logarithmic violations in $d=2$. Scaling properties of
various experimental observables are computed
in an expansion in $\epsilon=2-d$, and exactly in $d=1$.
\end{abstract}
\pacs{PACS: 75.10.Jm, 75.40.-s, 05.30.-d }
\narrowtext

\section{INTRODUCTION}
\label{intro}
Numerous experiments~\cite{expts} have by now
examined the properties of $S=1$
spin chain antiferromagnets which posses the Haldane gap~\cite{haldane}.
More recently, these antiferromagnets have been placed in
a strong magnetic field~\cite{fieldexpts},
and found to display evidence of a zero temperature
phase transition to a state with a non-zero ground state magnetization.
Theoretical studies~\cite{affleck,tsvelik,partha}
have also examined this
transition at zero temperature, and developed a
picture of it as the Bose condensation of
 magnons with azimuthal spin $S_z = 1$.
Although this possibility has not been considered before, it is not
difficult to see that the condensation of magnons in a finite
field should occur in quantum-disordered antiferromagnets in any
dimension $d$ (provided, of course, the magnons continue to behave as
bosons). In this paper we shall present a general theory, in
dimensions $d \leq 2$, of the
finite temperature properties of quantum antiferromagnets in the vicinity
of such a field-induced quantum transition.

We begin by elucidating the precise conditions under which our
results apply. Consider an antiferromagnet with a quantum
disordered ground state. The Hamiltonian must posses at least an
axial symmetry {\em i.e.\/} at least one component (say $z$)
of the total spin must commute with the Hamiltonian.
The lowest excitation with non-zero spin must be separated from
the ground state by a gap, and behave like a quasiparticle with
bosonic statistics. These conditions are clearly satisfied by Haldane
gap antiferromagnets in which the in-plane anisotropy can be
neglected (the compound NENP does have a rather small in-plane
anisotropy~\cite{expts},
and the restrictions this places on applying our results to
on experiments in NENP will be
discussed later). In $d=2$, the $S=1/2$ kagome Heisenberg
antiferromagnet~\cite{kagome} is
probably the most accessible candidate upon which our results can be
tested - it has been argued that the magnons in this system are
bosons~\cite{spn,kagsub}.

Now place this antiferromagnet in a magnetic
 field pointing along the direction of axial symmetry.
The eigenenergy, $\varepsilon (k)$
of single magnon (boson) quasiparticles with momentum $k$
in a  field $H$
then takes the form:
\begin{equation}
\varepsilon (k) = \Delta + \frac{\hbar^2 k^2}{2 m} - g \mu_B S_z H
\end{equation}
Here $\Delta$ is the magnon gap and $m$ is the quasiparticle mass,
both determined in the zero field antiferromagnet.
For the $S=1$ Haldane gap chain,
we have the azimuthal spin $S_z = 1$. In the kagome antiferromagnet,
the magnon excitations have been argued to be
spinons~\cite{spn,kagsub}: we therefore
expect $S_z = 1/2$. Later, we will discuss how the field induced
transition in the kagome antiferromagnet may offer a way of
experimentally/numerically determining the value of $S_z$.

The antiferromagnet will undergo a $T=0$ field-induced transition
at the field $H=H_c$ which is given exactly by
\begin{equation}
g \mu_B S_z H_c = \Delta
\end{equation}
In the vicinity of this transition, we may describe low energy
properties of the antiferromagnet by just studying an effective
Hamiltonian for the bosonic magnons~\cite{affleck}. The remainder of
this paper will therefore consider properties of the following
coherent state path integral
over the Bose field $\Psi (x, \tau)$
\begin{displaymath}
Z = \int {\cal D} \Psi \exp\left( -\frac{1}{\hbar} \int_0^{\hbar/k_B T}
d \tau {\cal L}(\tau) \right)
\end{displaymath}
\begin{eqnarray}
{\cal L}(\tau) = \int d^d &x& \left[
\hbar \Psi^{\ast} (x,\tau) \frac{\partial \Psi (x,\tau)}{\partial \tau}
- \frac{\hbar^2}{2m} \Psi^{\ast} (x,\tau)\nabla^2 \Psi (x,\tau) -
\mu |\Psi (x,\tau) |^2\right]\nonumber\\
&~&~~~~~~~~~~~~~~~~~~~~~~~~~~~~~~~~
+ \frac{1}{2} \int d^d x d^d x^{\prime} |\Psi (x,\tau) |^2 v(x-x^{\prime})
|\Psi (x^{\prime}, \tau )|^2
\label{coherent}
\end{eqnarray}
where $x$ is the $d$-component spatial co-ordinate, $\tau$ is the
Matsubara time, the chemical potential
\begin{equation}
\mu = g \mu_B S_z H - \Delta,
\label{muvalh}
\end{equation}
and $v$ is a repulsive interaction of a short range $\sim \Lambda^{-1}$.
This field theory has a phase transition exactly at $T=0$, $\mu=0$
which has been studied by Fisher {\em et. al.\/}~\cite{weich}.
They identified the upper-critical dimension as $d=2$, above which
the interaction $v$ is irrelevant. For $d < 2$, $v$ is relevant,
but the exponents were nevertheless found to have trivial values:
the dynamic exponent $z=2$, the correlation length exponent
$\nu=1/2$, and the field anomalous dimension $\eta
=0$. The triviality of the exponents is partially related to the fact that
the parameter tuning the system through the transition, $\mu$, couples
to a conserved quantity - the density of bosons;
any such transition~\cite{conscharge}  must
have $z \nu = 1$.
The structure of the $d < 2$, finite-$v$ fixed point is thus very
unusual: despite describing a non-trivial, interacting, critical
field theory, the exponents associated with all the relevant
directions away from this fixed point are trivial. In this paper,
we shall show that the interactions are crucial in determining the
finite temperature properties of the Bose gas near this fixed
point. The fixed-point interactions are needed to preserve
hyperscaling for $d <2$ and lead to highly non-trivial scaling
functions for the finite-temperature correlations.

Before we state our zero scale-factor universality hypothesis for
$Z$ in its most general form, it is helpful to consider one of its
simple consequences at $T=0$. Examine the ground state boson density
$n = \langle | \Psi (x,\tau) |^2 \rangle$ as a function of $\mu$
for small $\mu$. This problem was studied many years ago for the
$d=3$ hard-sphere Bose gas~\cite{yang} with the result
\begin{equation}
n = \left[ \frac{2m\mu}{\hbar^2} \frac{1}{8\pi a} + {\cal O}(\mu^2 )
\right]\theta(\mu)~~~~~~~~d=3,~T=0
\label{lee}
\end{equation}
where $a$ is the hard-sphere radius, and $\theta (x)$ is the unit
step function. Note that, in addition to its dependence on $m$ and
$\mu$, the boson density is sensitive to the nature of the boson-boson
interactions (measured by the hard-sphere radius $a$).
A different choice for the boson-boson repulsion would lead to
different result for $n$. The situation in
dimensions $d < 2$ is however strikingly different; one manifestation
of the zero scale-factor universality is that for small $\mu$
\begin{equation} n  = \left( \frac{2 m \mu}{\hbar^2} \right)^{d/2} {\cal
C}~ \theta (\mu)~~~~~d<2,~T=0
\label{magt0}
\end{equation}
where ${\cal C}$ is a {\em universal number\/} {\em i.e.\/} independent
of the details of the interactions between the bosons.
We will determine ${\cal C}$ in a $d=2-\epsilon$ expansion; its exact
value in $d=1$ is known~\cite{affleck}: ${\cal C} = 1/\pi$.
The universality of ${\cal C}$ is a direct consequence
of having a
finite-coupling fixed point describing the onset at $\mu=0$.
For $d > 2$, interactions are irrelevant, which leads to a violation
of hyperscaling, and a dependence of the density on the
nature of the microscopic interactions (as in the $a$ dependence
of (\ref{lee})). Further, $n$ will depend linearly on $\mu$ for all $d
> 2$.
Precisely in $d=2$, as we shall see below, (\ref{magt0}) is
violated only by
a logarithmic dependence on the microscopic interactions.

For $d < 2$, the combination of the presence of hyperscaling,
and the absence of any anomalous exponents in the leading critical
behavior, leads to remarkably universal finite temperature
properties. As in Ref.~\cite{CSY}, we may use finite-size scaling
to deduce scaling forms away from $\mu=0$, $T=0$. However the absence
of any anomalous dimensions ($z=2$, $\eta=0$, $\nu=1/2$)
means that the usual two scale-factor
universality~\cite{twoscale} is now reduced to a {\em
zero scale-factor universality.\/} described more precisely in the
following subsection.

\subsection{Zero scale-factor universality}
\label{introzerosc}
In simple terms, this universality is just the statement that all
response functions are universal functions of the {\em bare\/} coupling
constants $\mu$ and $m$. There are no non-universal amplitudes; the usual
case has two non-universal amplitudes, or scale-factors~\cite{twoscale}.
The universality can be stated more precisely
in terms  of the the boson Green's function
\begin{equation}
G(x, \tau) = \langle {\cal T}\Psi (x, \tau) \Psi^{\ast} (0,0) \rangle
\end{equation}
where ${\cal T}$ is the ordering symbol in imaginary time $\tau$.
After Fourier transformation as per
\begin{equation}
G (k, i\omega_n ) = \int d^d x \int_0^{\hbar/(k_B T)}
d \tau e^{-i(k \cdot x - \omega_n \tau)} G (x, \tau)
\end{equation}
this yields $G(k, i \omega_n)$ at the Matsubara
frequencies $\omega_n = 2 \pi n T / \hbar$, from which the retarded Green's
function
$G^R (k, \omega)$ can be obtained by analytic continuation to real
frequencies using
\begin{equation}
G^R (k, \omega ) = - G(k, i\omega_n = \omega ).
\end{equation}
The Lehman spectral representation of the Green's function
implies that
$-\omega \mbox{Im} G^{R} (k,  \omega ) > 0$, but does not constrain
$G^{R}$ to be an odd function of $\omega$.
Our central result is the zero scale-factor universality of $G^R$,
which is equivalent to the scaling form
\begin{equation}
G^{R} ( k, \omega ) = \frac{\hbar}{k_B T} {\cal A} \left(
\frac{\hbar \omega}{k_B T} , \frac{\hbar k}{\sqrt{2mk_B T}},
\frac{\mu}{k_B T} \right)
\label{scalegr}
\end{equation}
where ${\cal A}$ is a highly non-trivial, but
completely universal complex-valued function; naturally,
${\cal A}$ is independent of the nature of the boson-boson repulsion.
An important property of ${\cal A}$ is that it is
{\em analytic at all finite, real, values of all
three arguments.\/}
Similar scaling forms hold for other correlators of $\Psi$ - a
particularly instructive observable is the {\em local\/} Green's
function, $G^R_{\ell}$ \begin{equation}
G^R_{\ell} ( \omega ) = \int \frac{d^d k}{(2 \pi)^d} G^R ( k , \omega )
\label{defloc}
\end{equation}
If we take the imaginary part of this equation, it is expected
that the resulting on-shell contributions on the right-hand-side will
occur only at small momenta (determined by the frequency $\omega$),
and the momentum integral is ultraviolet convergent. We may
therefore deduce from
(\ref{scalegr}) the scaling form
\begin{equation}
\mbox{Im} G^R_{\ell} (\omega ) = -\frac{1}{\omega}
\left| \frac{m\omega}{\hbar} \right|^{d/2}
F \left( \frac{\hbar\omega}{k_B T}, \frac{\mu}{k_B T} \right)
\label{deff}
\end{equation}
where $F$ is a fully universal, dimensionless, positive function.
It is quite instructive to consider the limiting behavior of $F$ for
small and large frequencies. We expect that $G^{R}_{\ell}$ should be
analytic at $\omega=0$ at any finite $T$; this combined with the
positivity condition on the spectral weight noted above, implies
$\mbox{Im} G^R_{\ell} ( \omega )
\sim \omega$ for small $\omega$
at finite $T$. Therefore, from (\ref{deff}) the scaling function,
$F$,
must satisfy $F( \overline{\omega}, t ) \sim
|\overline{\omega}|^{2-d/2}$ at small $\overline{\omega}$ (we use here
and henceforth the notation $\overline{\omega} \equiv \hbar\omega/(k_B
T)$, and $t \equiv \mu/(k_B T)$); the coefficient of the
$|\overline{\omega}|^{2-d/2}$ term is quite difficult to determine, and
will be obtained in this paper only in a  special limit.
For large $\omega$, or short times, $G^R$ should display essentially
free particle behavior, as the dilute bosons have not had enough time
to interact with each other.
Using the free boson spectral weight we can deduce that
\begin{equation}
F ( \overline{\omega} , t ) = \frac{(2 \pi)^{1-d/2}}{\Gamma(d/2)}
\theta( \omega)~~~\mbox{as $|\overline{\omega}| \rightarrow \infty$}
\label{freeparticles}
\end{equation}

Let
us conclude this subsection by noting the  precise conditions under which
the system is in the critical region and (\ref{scalegr}) and
(\ref{deff}) are valid. We must have
\begin{equation} |\mu|, k_B T \ll
\frac{\hbar^2 \Lambda^2}{2m}, v(0)
\label{condmu}
\end{equation}
Further the measurement wavevectors must satisfy
\begin{equation}
k \ll \Lambda
\label{condk}
\end{equation}

In $d=2$ the zero-scale-factor universality is violated by a
logarithmic dependence on the microscopic interactions. Furthermore,
the scaling function ${\cal A}$ will have a singularity associated with the
finite temperature Kosterlitz-Thouless transition. Zero scale-factor
universality does not hold for $d>2$.

\subsection{Neutron scattering}
The dynamic information contained in $G^{R}$ is directly observable in
neutron scattering experiments.
In Appendix~\ref{shankar} we discuss the relationship between the correlators
of $\Psi$ and antiferromagnetic correlations measured by the neutrons; this
discussion is limited to the case where the quantum disordered phase has
confined spinons {\em i.e.\/} $S_z = 1$, as in Haldane gap antiferromagnets.
The relationship for the case of deconfined spinons ($S_z = 1/2$) is quite
different~\cite{triangle} and will not be considered in this paper explicitly.

In the following,  $\hbar \omega$ will measure the energy {\em lost\/} by the
neutrons in their interaction with the antiferromagnet. Consider first a
scattering even in which the antiferromagnet undergoes a $\Delta S_z = +1$
transition. Then, from the discussion in Appendix~\ref{shankar} and the
fluctuation-dissipation theorem we may conclude that (for the scattering
cross-section is given by the dynamic structure factor $S_{+-} ( k, \omega )$
where  \begin{equation} S_{+-} ( k,
\omega ) = \frac{ - 2 Z \mbox{Im} G^{R} ( k, \omega )}{ 1 -
e^{-\hbar\omega/k_B T}} \label{introz} \end{equation}
where $Z$ is
a {\em non-universal\/} quasi-particle renormalization
factor between the magnon operators and  the ones that couple to the
neutrons. (The wavevector $k$ on the left-hand-side is measured from the
antiferromagnetic ordering wavevector)
 The scaling result
for $S_{+-}$ thus follows directly from (\ref{scalegr}).
Next, consider scattering with $\Delta S_z = -1$ for the
antiferromagnet.
The associated dynamic structure factor is then $S_{-+} ( k, \omega )$
which is
\begin{equation}
S_{-+} ( k, \omega ) = \frac{ 2 Z \mbox{Im} G^{R} ( k, -\omega )}{
1 - e^{-\hbar\omega/k_B T}}
\end{equation}

By not resolving the energy of the scattered neutron, it is also
possible to measure equal-time correlations embodied in the static
structure factors.  However the scaling properties of these
observables are subtle, and require more careful interpretation.
By performing a weighted frequency integral over the scaling limit
of $G^R$, we are implicitly only sensitive to frequencies much smaller
than a high frequency cutoff like $\Delta/2$. Thus in the following,
our equal-time structure factors actually refer to scattering
experiments in which energy transfers greater than $\Delta/2$ are
not integrated over. Not blocking out these events, will produce a
background structure factor which may (as in the Haldane gap region
defined below) overwhelm the universal part of the static structure
factor which is considered here. Keeping this caveat in mind, we define
the structure factor $S_{+-} (k)$ \begin{equation}
S_{+-} ( k) = \int_{\sim-\Delta/2}^{\sim \Delta/2} \frac{d \omega}{2
\pi} S_{+-} ( k , \omega )
\label{chipm}
\end{equation}
where the result is not sensitive to the precise locations of the
limits. The correlator $S_{-+}$ can be defined analogously.
{}From
(\ref{chipm}) and (\ref{scalegr}) we can deduce the
scaling result
\begin{equation}
S_{+-} (k) = Z {\cal
B}_{+-} \left( \frac{\hbar k}{\sqrt{2mk_B T}}, \frac{\mu}{k_B T} \right)
\label{defb1}
\end{equation}
where ${\cal B}_{+-}$ is a universal scaling function determined
completely by ${\cal A}$ (similarly for $S_{-+} (k)$).
Finally, using the fact that $\Psi$ and $\Psi^{\ast}$ are canonically
conjugate fields, it possible to deduce a frequency sum-rule on
$\mbox{Im} G^{R}$ which leads to
\begin{equation}
{\cal B}_{+-} (r, t) = {\cal B}_{-+} (r,t) + 1
\label{bb1}
\end{equation}
where we will henceforth use $r \equiv \hbar k /\sqrt{2 m k_B T}$.

We reiterate that all of the above results in this subsection refer only to
the case of antiferromagnets with confined spinons.

\subsection{Uniform magnetization}
A second useful set of observables are those
associated with magnetic fluctuations around $k=0$. The uniform magnetization
is a conserved quantity and, as a result, the quasi-particle renormalization
factor $Z$ does not appear in their scaling forms. The simplest of these is
mean-value of the
magnetization density $M$ itself, which is of course related to
magnon density by
\begin{equation}
M = g \mu_B S_z n.
\label{mvaln}
\end{equation}
This relationship, and the considerations of this subsection, apply to both
confined ($S_z=1$) and deconfined ($S_z=1/2$) spinons.
The mean-value of the magnetization is therefore
\begin{equation}
M = g \mu_B S_z G(x=0, \tau=0^{-})
\end{equation}
and obeys the scaling form
\begin{equation}
M = g \mu_B S_z \left( \frac{2 m k_B T}{\hbar^2} \right)^{d/2}
{\cal M} \left ( \frac{\mu}{k_B T} \right)
\label{defb2}
\end{equation}
where the function ${\cal M}$ is again dependent only on ${\cal
A}$. The result (\ref{magt0}) follows from (\ref{defb2}).

\subsection{Phase diagram}
The above discussion shows that the scaling functions of a large number
of experimental observables can be obtained directly from the
primary scaling function ${\cal A}$. The
remainder of the paper is therefore devoted primarily to describing
${\cal A}$ and associated scaling functions in different parameter
regimes.
It is convenient to discuss the properties of ${\cal A}$ separately in
three distinct regimes, which are analogous to those
found by Chakravarty {\em et. al.\/}~\cite{CHN} in the $d=2$ $O(3)$
sigma model. These regimes are illustrated in the phase
diagrams in Figs~\ref{figd1}
($d=1$) and~\ref{figd2} ($d=2$). The crossover boundaries between the
regimes are delineated by the value of the dimensionless ratio
$\mu / k_B T$ (upto logarithmic terms in $d=2$). We
consider the three cases separately

\subsubsection{$\mu \ll - k_B T$}
This is the analog of the quantum-disordered regime of Refs~\cite{CHN,CSY}.
Only a dilute gas of thermally-excited bosons is present, and their
mutual interactions are weak. Properties of the quantum
antiferromagnet can be described by a low-magnon-density expansion
about the quantum disordered ground state.
In $d=1$, we identify this as the Haldane-gap regime
in Fig~\ref{figd1}.

\subsubsection{$|\mu| \ll k_B T$}
This is the quantum-critical regime, in which $z=2$ critical
fluctuations are quenched in a universal way by the temperature.
The value $k_B T$ is the dominant energy scale and universally
determines everything: the boson density, spectrum, and interactions.
The only small parameter which may be used to determine scaling
functions is $\epsilon=d-2$. As we will see, this is not particularly
effective in $d=1$ where, fortunately, exact methods are
available.

It is instructive to consider the physics of this region as a function
of the measurement frequency $\omega$ (see Fig~\ref{omega}).
At large frequencies, $\omega \gg k_B T$, or short times, the effects
finite temperature have not yet become manifest, and the system
displays the physics of the $\mu=T=0$ critical field theory {\i.e.\/}
it is a dilute gas of bosonic quasiparticles with repulsive
interactions. As one lowers the frequency through $k_B T/\hbar$ there
is a crossover to a novel $z=2$, quantum-relaxational regime
(Fig~\ref{omega}). Now each boson interacts strongly with thermally
excited partners, leading to strong dissipation and overdamped
quasiparticles.

\subsubsection{$\mu \gg k_B T$}
The behavior in $d=1$ and $d=2$ is quite different and we will
therefore consider the two cases separately.

In $d=1$, the ground state
is a Luttinger liquid, which is itself a critical phase with $z=1$
(Fig~\ref{figd1}).
Again, consider the physics as a function of $\omega$ (Fig~\ref{omega}).
For sufficiently large $\omega$ ($\hbar\omega \gg \mu$) we have the
dilute bose gas physics of the $\mu=T=0$ critical point, similar to that
discussed above for the quantum-critical region. At smaller
$\omega$, there is a crossover (near $\hbar \omega = \mu$) to a
Luttinger liquid-like region where we may as well assume that $T=0$.
However at small enough frequencies, $\hbar \omega \sim k_B T$,
the effects of a finite temperature finally become apparent.
The massless modes of the Luttinger liquid are then quenched into a $z=1$
quantum-relaxational regime, rather similar to the $z=2$ quantum
relaxational regime discussed above. This last crossover is, strictly
speaking, a property of the $z=1$ critical point on the Luttinger liquid
fixed line determined by the value of $\mu$, and not a property of the
$z=2$ critical end-point at $\mu=T=0$. It is thus described by a reduced
Luttinger liquid scaling function ${\cal A}_L$. We will obtain  exact
results for ${\cal A}_L$ using an argument based on conformal
invariance. We will also discuss an important compatibility condition
between the scaling functions ${\cal A}$ and ${\cal A}_L$, and show
how the more general ${\cal A}$ collapses into the small $\mu$ limit of
${\cal A}_L$.

In $d=2$, (Fig~\ref{figd2})
the ground state is a boson superfluid, which survives
at finite temperature. There is a Kosterlitz-Thouless phase
transition at a finite $T$, which has been studied
in some detail earlier~\cite{popov,fisher}. In the superfluid phase
therefore, one has a large frequency dilute Bose gas behavior crossing
over to a small frequency Goldstone phase with quasi-long-range order
(Fig~\ref{omega}).

The outline of the rest of the paper is as follows.
We will begin, in Section~\ref{rganal}, by a renormalization group
analysis of the partition function $Z$.
This will allow us to demonstrate the logarithmic corrections to
scaling in $d=2$ and obtain the leading terms of the
in a $\epsilon=2-d$ expansion. These leading terms
are consistent with the zero scale-factor universality.
We will then, in Section~\ref{onedimension}, turn to a discussion of the
exact properties of  ${\cal A}$ and ${\cal A}_L$ in $d=1$.
A brief summary and a discussion of relevance to experiments appear
in Section~\ref{conc}. Three appendixes contain discussions
of some peripheral points.

\section{RENORMALIZATION GROUP ANALYSIS}
\label{rganal}
The momentum shell renormalization group equations for $Z$ have
already been obtained by Fisher and Hohenberg~\cite{fisher}. However
their analysis of the equations was restricted to $d=2$, $\mu > 0$, and
temperatures at or below the Kosterlitz Thouless transition. In this
section we will extend this analysis to cover the remaining regions
in $d=2$ (Fig~\ref{figd2}), and to dimensions $d<2$ in an
$\epsilon=2-d$ expansion (Fig~\ref{figd1}).
The analysis~\cite{fisher} proceeds by introducing
an upper cutoff, $\Lambda$ in momentum space, and replacing
$v$ by a contact interaction $u=\Lambda^{-d} v(0)$.
Degrees of freedom in a shell between
$\Lambda$ and $\Lambda e^{-\ell}$ are integrated out, followed by a
rescaling of co-ordinates and field variables
\begin{eqnarray}
x^{\prime} = e^{-\ell} x ~~~&~&~~~\tau^{\prime} = e^{-2 \ell} \tau
\\
\Psi^{\prime} (x^{\prime}, \tau^{\prime} ) &=& e^{d\ell/2} \Psi ( x,\tau )
\nonumber
\end{eqnarray}
Note that the there is no anomalous dimension in the rescaling
factor for $\Psi$ as $\eta=0$. Further, the scaling dimension of
$|\Psi |^2$ is exactly $d$, as it must be for any conserved charge
density~\cite{conscharge}.

It is convenient to consider the cases $T=0$ and $T>0$ separately:

\subsection{$T=0$}
The renormalization group equations are~\cite{fisher,weich}
\begin{eqnarray}
\frac{d \mu}{d\ell} &=& 2 \mu\\
\frac{d u}{d \ell} &=& (2-d) u -
\frac{mK_d \Lambda^{d-2}}{\hbar^2 } u^2
\end{eqnarray}
where $K_d = S_d / (2 \pi)^d$ and $S_d$ is the surface area of a unit sphere
in $d$ dimensions. We integrate these equations to a scale
$e^{\ell^{\ast}}$ where the system is non-critical {\em i.e.}
when $\mu \sim \alpha \hbar^2 \Lambda^2 / (2m)$ where $\alpha$ is
significantly smaller than unity, but not so small that the system is
still critical.
For $\mu > 0$,
the magnetization, $M$, and the boson density $n= M/(g \mu_B S_z)$
are then given by
\begin{equation}
n =  e^{-d \ell^{\ast}} \frac{\mu (\ell^{\ast})}{u (
\ell^{\ast})}
\label{mrg}
\end{equation}
For $d < 2$, and $\epsilon = 2-d$ small, $u$ approaches its fixed
point value
\begin{equation}
u^{\ast} = \frac{\hbar^2}{K_d \Lambda^{d-2} m} \epsilon
\label{ufixed}
\end{equation}
Inserting this fixed-point value and the dependence of $\ell^{\ast}$
on the initial value of $\mu$ into (\ref{mrg}) we find to lowest
order in $\epsilon$ that $n$ indeed has the form (\ref{magt0}) with
the universal number ${\cal C}$ given by
\begin{equation}
{\cal C} = \frac{1}{4 \pi \epsilon}
\end{equation}

In $d=2$, $u$ approaches 0 logarithmically slowly. For
large ${\ell}$ we have
\begin{equation}
u(l) = \frac{ 2 \pi \hbar^2}{m \ell}
\label{ud2}
\end{equation}
Inserting this into (\ref{mrg}) we find~\cite{popov}
\begin{equation}
n = \frac{m \mu}{4 \pi \hbar^2} \log \left(\frac{\hbar^2
\Lambda^2}{2m \mu} \right)
\label{popovres}
\end{equation}
Note the logarithmic violation of the perfect scaling of (\ref{magt0}).

\subsection{$T > 0$}
We will restrict our analysis to the center of the quantum critical
region in $d=2$ (Fig~\ref{figd2}) and $d<2$ (Fig~\ref{figd1}):
the initial value of $\mu$ will therefore be fixed at $\mu ( \ell =
0)=0$ and the initial value of the temperature $T$ will be close to 0.
We will only need the finite $T$ renormalization group equations for
$\mu$ and $T$ which are~\cite{fisher}
\begin{eqnarray}
\frac{d T}{d\ell} &=& 2 T \nonumber \\
\frac{d \mu}{d \ell} &=& 2 \mu - \frac{2 \Lambda^d K_d
u}{\displaystyle \exp \left( \frac{1}{k_B T} \left( \frac{\hbar^2
\Lambda^2}{ 2 m} - \mu\right)\right) -1 }
\label{rgt}
\end{eqnarray}
To leading order
in $\epsilon$ (leading-logs) it is sufficient to assume that
for $d < 2$ ($d=2$) $u$ is
given by Eqn (\ref{ufixed}) (Eqn (\ref{ud2})).
We will now integrate the renormalization group equations until
a scale $\ell^{\ast}$ where
\begin{equation}
\mu ( \ell^{\ast}) = - \alpha\frac{\hbar^2 \Lambda^2}{2 m}
\label{deflst}
\end{equation}
The correlation length $\xi$ is then given by
\begin{equation}
\xi = \frac{e^{\ell^{\ast}}}{\Lambda\sqrt{\alpha}}
\label{defxi}
\end{equation}
while the boson density is
\begin{eqnarray}
n &=& e^{-d \ell^{\ast}} \langle | \Psi_{\ell^{\ast}} ( x=0,
\tau=0^{-}) |^2 \rangle \nonumber \\
&=&  e^{-d \ell^{\ast}} \int \frac{d^d k}{(2 \pi)^d}
\frac{1}{ \displaystyle \exp \left( \frac{1}{k_B T (\ell^{\ast})}
\left( \frac{\hbar^2
k^2}{ 2 m} - \mu ( \ell^{\ast}) \right)\right) -1 } \nonumber\\
&\approx& - \frac{e^{- d \ell^{\ast}}}{4 \pi} \left( \frac{2 m k_B T
(\ell^{\ast})}{\hbar^2} \right)^{d/2}
\log( 1 - e^{\mu(\ell^{\ast})/T(\ell^{\ast})})
\label{resn}
\end{eqnarray}
In the last step we have anticipated that to leading order in
$\epsilon$ it is sufficient to evaluate the integral directly in $d=2$.

Let us now examine the results of integrating (\ref{rgt}) for $d <
2$. To leading order in $\epsilon$ we find
\begin{equation}
\mu( \ell ) = - 4 \epsilon e^{2 \ell} \frac{\hbar^2 \Lambda^2}{2m}
\int_0^{\ell} \frac{ e^{-2 \ell^{\prime}} d \ell^{\prime}}{
\displaystyle \exp\left( \frac{ \hbar^2 \Lambda^2}{2 m
T(\ell^{\prime})} \right) - 1}
\label{muint}
\end{equation}
Using $T(\ell) = T e^{2 \ell}$, it is straightforward to perform the
integration and obtain from (\ref{deflst}) the leading result for
${\ell^{\ast}}$:
\begin{equation}
e^{-2 \ell^{\ast}} = \frac{2 m k_B T}{\hbar^2 \Lambda^2} 2 \epsilon
\log \left( \frac{1}{2 \epsilon} \right).
\end{equation}
{}From (\ref{defxi}) we therefore deduce
\begin{equation}
\xi = \frac{1}{(2 \epsilon \log(1/2 \epsilon))^{1/2}}
\frac{\hbar}{\sqrt{2 m k_B T}}
\end{equation}
and from (\ref{resn}) we obtain for the boson density
\begin{equation}
n = \left( \frac{2 m k_B T}{\hbar^2} \right)^{d/2} \frac{1}{4 \pi}
\log \left( \frac{1}{2 \epsilon \log (1/2 \epsilon)} \right)
\end{equation}
These last two results are consistent with the zero-scale-factor
universality of (\ref{defb1}) and (\ref{defb2}) and yield properties
of the scaling functions ${\cal B}_{+-}$ and ${\cal M}$ at $\mu=0$.

Finally consider properties in the quantum-critical region
in $d=2$.  The analog of Eqn (\ref{muint}) is
\begin{equation}
\mu( \ell ) = - 4 e^{2 \ell} \frac{\hbar^2 \Lambda^2}{2m}
\int_0^{\ell} \frac{d \ell^{\prime}}{\ell^{\prime}}
\frac{ e^{-2 \ell^{\prime}}}{
\displaystyle \exp\left( \frac{ \hbar^2 \Lambda^2}{2 m
T(\ell^{\prime})} \right) - 1}
\end{equation}
Integrating this to leading-log accuracy and using (\ref{deflst}) we
find
\begin{equation}
e^{-2 \ell^{\ast}} = 4 \frac{2 m k_B T}{\hbar^2 \Lambda^2 \alpha}
\log \log \left( \frac{\hbar^2 \Lambda^2}{2 m k_B T} \right)
\end{equation}
We therefore have from (\ref{defxi}) for the correlation length
\begin{equation}
\xi = \frac{\hbar}{\sqrt{2 m k_B T}} \frac{1}{\Bigl[4 \log \log (\hbar^2
\Lambda^2 / (2 m k_B T))\Bigr]^{1/2}}
\end{equation}
which violates the universality of $\ref{defb1}$ at $\mu=0$
by the double logarithms.
{}From (\ref{resn}) we get for the boson density
\begin{equation}
n = \frac{2 m k_B T}{\hbar^2} \frac{1}{4 \pi} \frac{1}{\Bigl[ \log ( \hbar^2
\Lambda^2 / ( 2 m k_B T))\Bigr]^4},
\end{equation}
again logarithmically violating (\ref{defb2}) at $\mu=0$.

\section{EXACT RESULTS IN ONE DIMENSION}
\label{onedimension}
We have so far determined that for small $\epsilon=2-d$ the
$\mu=T=0$ critical field theory has a contact interaction of strength
$u^{\ast} =  {\cal O}(\epsilon)$, and all other two/multi-particle
interactions can be neglected. Remarkably, following
Haldane~\cite{luttinger}, it also possible to  determine the exact
critical field theory for $\epsilon=1$ or $d=1$. The critical field
theory then has $u^{\ast}|_{\epsilon=1}  = \infty$ (the bosons are thus
impenetrable). Moreover, all other boson interactions continue to be
irrelevant. In Appendix~\ref{canonical} we consider a one-dimensional
Bose gas in the vicinity of this strong-coupling fixed point and
demonstrate this explicitly.

The methods of Appendix~\ref{canonical} and earlier
works~\cite{luttinger,1dbose,affleck} use the well-known equivalence between
the $d=1$ impenetrable Bose gas and free fermions. The field theory
of the critical end-point at $\mu=T=0$ is therefore given
by the
free fermion Hamiltonian
\begin{equation}
H_F = \int dx \Psi_F^{\dagger} (x) \left(
-\frac{\hbar^2}{2m} \frac{d^2}{dx^2} - \mu \right) \Psi_F (x)
\end{equation}
where $\Psi_F$ are canonical fermion fields. Correlators of this theory
can only depend upon $m$, $\mu$, and $T$, and the zero scale-factor
universality is therefore manifest. The scaling limit of the correlators
of the uniform magnetization can now be obtained almost
trivially: the uniform magnetization density just measures the
number of particles and its correlators are therefore the same as  those
of $g \mu_B S_z \Psi_F^{\dagger} \Psi_F$.
In particular we have
for the scaling function for the uniform magnetization in (\ref{defb2})
\begin{equation}
{\cal M} ( t) = \frac{1}{\pi} \int_0^{\infty} dy \frac{1}{
e^{y^2 - t} + 1}
\end{equation}
where $t\equiv\mu/(k_B T)$.
This scaling function has the limiting value
${\cal M} = e^{t} /(2 \sqrt{\pi})$ as $t \rightarrow -\infty$ in
the Haldane gap region, and ${\cal M} = \sqrt{t}/\pi$ as $t
\rightarrow \infty$ in the Luttinger liquid region (Fig~\ref{figd1}). This
last result combines with (\ref{defb2}) to yield (\ref{magt0})
with ${\cal C} = 1/\pi$~\cite{affleck}.

Observables associated with correlations of the staggered
magnetization, like $G^{R}$,
are much more difficult to obtain - it is necessary to express the
impenetrable Bose fields in terms of the Fermi fields by a continuum
Jordan-Wigner transformation~\cite{luttinger} and then evaluate the
correlator - a naive Wick's theorem expansion of this correlator will
yield an infinite number of terms.
Recently, Korepin and Slavnov~\cite{korep3}, following earlier work of
Lenard~\cite{lenard}, have succeeded in resumming this expansion and
showing that all space-time dependent, finite temperature, correlators of the
impenetrable Bose gas can be expressed in terms of the solution and Fredholm
determinant of a linear Fredholm integral equation. Thus determination
of the universal scaling function ${\cal A}$ in $d=1$ has been reduced
to the problem of solving completely an integral equation, and taking
the Fourier transform of the result.
Analytic methods can take us no further, and it is necessary to resort
to numerical analysis of the integral equations. We have begun such a
numerical program, and have so far obtained essentially exact results
for the equal-time correlations - these are described below in
Section~\ref{eqtime}. It should
be possible to extend our results to obtain local, time-dependent
correlations (and hence the scaling function $F$ in (\ref{deff})) but we
have not yet done so. A general picture of the form of $F$ can be
obtained from the asymptotic limits quoted in
Section~\ref{introzerosc}; should it become experimentally useful to
obtain more precise numerical results for $F$, we shall be happy to
provide them.

We also note that, recently,
Korepin and collaborators~\cite{korep1,korep2} have succeeding in
determining exact results for certain asymptotic properties of $G^{R}$
by applying the quantum-inverse scattering method to the integral
equations noted above. For the equal-time $G^{R}$ they obtained results
for the leading and next-to-leading terms as $x \rightarrow \infty$,
while for unequal time correlators, both $x$ and $\tau$ were sent to
$\infty$. Unfortunately, these asymptotic results are not very useful in
determining experimental observables which require Fourier
transformation to  functions of momenta and frequency (the large $x$
behavior of a function implies little about the small $k$ limit of its
Fourier transform). Simply Fourier transforming the asymptotic terms
leads to results which compare very poorly with the exact results which
we obtained by the alternative means described below. We comment on some
features of these exact asymptotic results in Appendix~\ref{itsetal}.

In Section~\ref{lutliq} we
will consider the limit $\mu \gg k_B T$ where it is possible to make
much greater analytic progress in determining the scaling functions.
As
we have already noted,
the lower frequency properties in this region are
described by Luttinger liquid
criticality, and it possible to use conformal invariance arguments to
obtain closed-from results.

\subsection{Equal-time Structure Factor}
\label{eqtime}

The most convenient procedure for determining equal-time correlations
begins with Lenard's~\cite{lenard} result for the density matrix
of the impenetrable Bose gas, which is tantamount to a formal solution
of the integral equation of Ref~\cite{korep3}. His result can be written
as~\cite{korep3}
\begin{equation}
G (x, \tau=0^{-}) = \langle 0 | \hat{G}_F
(1 - 2 \hat{G}_F )^{-1} | x \rangle
\mbox{det} ( 1 - 2 \hat{G}_F )
\label{lenard}
\end{equation}
where the operator $\hat{G}_F$ acts on the
real axis between 0 and $x$, and has the matrix elements
\begin{equation}
\langle x | \hat{G}_F | x^{\prime} \rangle =
\int_{-\infty}^{\infty} \frac{dk}{2 \pi} \frac{e^{ik(x-x^{\prime})}}{
e^{(\hbar^2 k^2 / (2m) - \mu)/(k_B T)} + 1},
\end{equation}
{\em i.e.\/} the fermion Green's function.
The form (\ref{lenard}) is amenable to rapid numerical evaluation.
We discretize the real line between 0 and $x$ into $N$ points, whence
the operators in (\ref{lenard}) become $N\times N$ Toeplitz matrices.
A straightforward extension of
Levinson's algorithm~\cite{levinson} was then used to compute
the determinant and inverse of $1 - 2 \hat{G}_F$. The computer time
required for this step scales only as $N^2$, and we were able to
easily
uses values upto $N=10000$. The results for $G$ for large $x$
were compared with the exact asymptotic results of Ref~\cite{korep1},
with excellent agreement. Finally, we performed a spatial Fourier
transform, and obtained results for the scaling function ${\cal B}_{-+}$
of the structure factor, $S_{-+}(k)$ defined below (\ref{defb1})
and by (\ref{bb1}).

Our results for some representative values of $t = \mu / (k_B T)$ are shown
in Fig~\ref{figb1}. A computer program to obtain numerical
values of ${\cal B}_{-+} (r, t)$ ($r = \hbar k /\sqrt{2mk_B T}$)
for arbitrary $r,t$ can be obtained from the authors; the accuracy is limited
only by computer time, but it is possible to obtain 3 significant figure
accuracy quite rapidly.

We note that the zero temperature limit of $S_{-+}(k)$ was
computed by Vaidya and Tracy~\cite{vaidya}: they also pointed out
that the $T=0$, $S_{-+}(k)$ has non-analyticities~\cite{luttinger}
(which are, however, unobservably
weak) at integer multiples of $2 k_F$, where $k_F$ is the Fermi wavevector
of $H_F$

\subsection{Luttinger Liquid}
\label{lutliq}
This is the regime where $\mu \gg k_B T$. In the fermionic description of the
problem, this means the Fermi sea is much deeper than the temperature.
In this case the ground state and its pertinent excitations can be
described by a theory in which the spectrum is linearized near the Fermi
points $k = \pm  k_F$ determined by
\begin{equation}
\mu ={ \hbar^2 k^{2}_{F} \over 2m}.
\end{equation}
By the same token, as long as the probe frequency $\omega$ and momentum
$k$ do not cause excitations that probe the sea deeply (i.e. stay in the
linearized region)
the Bose gas will  exhibit characteristics of the line of finite-$\mu$,
$z=1$ Luttinger-liquid critical-points at $T=0$ - the
Luttinger-liquid scaling function for the Green's function will be
denoted by ${\cal A}_L$. In terms of Fig~\ref{omega}, it means that the
${\cal A}_L$ will describe the lower frequency crossover
around $\hbar \omega \sim k_B T$. The ratio $\hbar\omega/k_B T$ can take
arbitrary values as long as both $\hbar\omega$ and $k_B T$ remain
significantly  smaller than $\mu$. The depth of the Fermi sea will not
enter any of the calculations of ${\cal A}_L$ and $\mu$
will enter only via the Fermi velocity $c$. For example, in the low
density limit, $c$ is given by
\begin{equation}
c = {k_F \over m} =
\left({2 \mu \over m}\right)^{1/2}.
\end{equation}
We shall see all this happen as we analyze the exact
results momentarily.
The crossover around $\hbar \omega \sim \mu$ in Fig~\ref{omega} is
{\em not\/} part of the Luttinger liquid criticality, and is instead
associated with the $z=2$ critical end-point, and the scaling function
${\cal A}$. It should be apparent from this discussion that the limits
$\hbar \omega /k_B T \rightarrow \infty$ and $\mu/k_B T \rightarrow
\infty$ of ${\cal A}$ do {\em not\/} commute.

Note that Luttinger-liquid
criticality, and associated scaling forms, hold even
when the condition that $\mu$ be small
in (\ref{condmu}) is violated.
Suppose, however, that $k_B T,\hbar \omega$  and $\hbar^2 k^2 /(2m)$ are
much smaller than $\mu$, and that $\mu $ itself is small so that
Eqn.(\ref{condmu}) is satisfied. {\em Then zero scale-factor
universality of the $z=2$ critical point at $\mu=0$, $T=0$ must be
simultaneously satisfied.} This will lead to a compatibility condition
condition between the scaling functions ${\cal A}$ and ${\cal A}_L$,
which
we shall shortly examine.

Let us begin by writing  down the scaling forms of the Luttinger liquid
at $T=0$.
We know that at equal times
\begin{equation}
G(x, \tau=0^{-} ) = \frac{D}{x^{\eta}}~~~;~~~x \rightarrow \infty,~T=0,
\mu > 0.
\label{defeta}
\end{equation}
The constant $D$, and the exponent $\eta$, will, in general,
 have a non-universal dependence
upon the microscopic couplings. However, knowledge of $D$, $\eta$,
and a zero-sound velocity, $c$, will universally determine
all remaining hydrodynamic properties in the Luttinger liquid regime. For
example the unequal time correlation function will have exactly the same
form as above with $x$ replaced by the euclidean distance $\sqrt{x^2 +
c^2 \tau^2}$. As for $\eta$, it has a value that depends on the Luttinger
coupling. At the point $\mu = 0$,
the bosons are impenetrable and equivalent to free fermions. For $\mu >
0$, the deviation from the impenetrability condition can be translated
into a residual interaction between fermions by integrating out doubly
occupied states (see Appendix~\ref{canonical}). This is the marginal
coupling of the Luttinger liquid. Let us note for future reference that
 $\eta =1/2$ for zero Luttinger coupling. Some readers may have trouble
reconciling this with the fact  that fermion-fermion correlation
functions fall as $1/x$ in free-field theory.
However, we have already noted that there is a rather complicated
between the Fermi and Bose fields and
we remind the reader of the
chain of transformations relating the two. First the hard-core bosons
are described by the Pauli matrices $\sigma_{\pm}$. The latter are than
converted to a single component fermions $\Psi_F$  by a Jordan-Wigner
transformation. These fermions are filled upto some Fermi momentum
determined by the chemical potential. When linearized near the Fermi
points, the spinless Fermi field turns into a pair of relativistic
fields $\Psi_{L,R}$. The hard-core boson correlation function at equal
times is
\begin{equation}
<\Psi (0) \Psi^{\dag} (x)> = <\Psi_F (0) e^{i
\pi \int_{0}^{x} \Psi_F^{\dag}(x') \Psi_F (x') dx'}\Psi_F^{\dag}(x)>.
\end{equation}
If we now write
\begin{equation}
\Psi_F(x) = \Psi_{L} e^{-ik_F x} + \Psi_R e^{ik_F x}
 \end{equation}
and drop terms that oscillate at $k_f$, we obtain
\begin{equation}
<\Psi (0) \Psi^{\dag} (x)> \simeq < \Psi_L (o) e^{i \pi
\int_{0}^{x}(\Psi_{L}^{\dag}(x') \Psi_L (x') + \Psi_{R}^{\dag}(x')
\Psi_R (x'))dx'} \Psi_L^{\dag}(x) + L \leftrightarrow R>.
\end{equation}
Clearly this is a complicated object in the Fermi theory. To evaluate it
one uses bosonization.  Using the standard dictionary
it is possible to show that it is proportional to the two-point function
\begin{equation}
<e^{-i \sqrt{\pi}\tilde{\phi} (0)}e^{i \sqrt{\pi}\tilde{\phi} (x)}>
\propto {1 \over x^{1/2}}
\end{equation}
where $\tilde{\phi}$ is the field dual to the usual boson field. (See for
example Ref~\cite{shankbozo}).

We now consider the correlations at finite temperatures. In general the passage
from zero to nonzero temperatures is nontrivial since in the latter case non
only the ground state but excited state correlations enter. However in two
euclidean dimensions, in a relativistically invariant theory such as this one,
we have the remarkable result from conformal field theory that was first
pointed out by Cardy ~\cite{cardy}.  Using the conformal mapping between the
infinite-plane and the strip
of finite-length $L_{\tau} = \hbar c / k_B T$ along the imaginary-time
direction, one can obtain from eqn.(\ref{defeta})
\begin{equation}
G_L(x, \tau) = \frac{D}{2^{\eta /2}} \left( \frac{2 \pi k_B T}{\hbar c}
\right)^{\eta} \left[\cosh\left(\frac{2\pi k_B T x}{\hbar c}\right)
- \cos \left(\frac{2 \pi k_B T \tau}{\hbar}\right)\right]^{-\eta /2}
\label{cardy}
\end{equation}
(The operator  $e^{-i \sqrt{\pi}\tilde{\phi} (0)}$ is a primary field, which
allows us use conformal invariance as above.)
The subscript $L$ has been placed to emphasize that this formula is
valid only near the lower frequency crossover in Fig.~\ref{omega}

The ease with we obtain this result should not detract us from noting its
importance-- rarely does one have the thermally averaged correlation
functions of an interacting system.
We shall therefore spend some time analyzing this result.

As a first step let us extract form this the correlation function at
equal time, for long distances. It is readily seen that
\begin{equation}
G(x,0^{-}) = D \left( \frac{2 \pi k_B T}{\hbar c}
\right)^{\eta} \exp \left(\frac{-\eta \pi k_B T x}{\hbar
c}\right)~~~;~~~ T>0,~\mu > 0,~ x \rightarrow \infty
\label{luttcorr}
\end{equation}

Now let us consider the regime where (\ref{condmu}), is also satisfied
and so zero-scale-factor universality holds.
Thus $D$, $\eta$, and $c$ can no longer be nonuniversal, but must be
universal functions of $\mu$ and $m$. Let us now invoke the asymptotic
(long distance) hard-core boson scaling functions at the $\mu =0$
critical point which are known by exact solution~\cite{vaidya,korep1}:
\begin{equation}
G(x,0^{-}) = {\sqrt{2mk_BT} \over \hbar}{\rho_{\infty}\over
\sqrt{\pi}}exp\left( -
\frac{\sqrt{2 m k_B T}}{\hbar }{ \pi x \over 4 \sqrt{t}}\right)
\end{equation}
where $t= \mu /(k_BT)$. The constant
$\rho_{\infty}$ is a known universal number; more details on  this
formula are relegated to Appendix~\ref{itsetal}.
We see that this agrees with the Luttinger liquid result
Eqn.(\ref{luttcorr}) if
\begin{equation}
\eta=\frac{1}{2}~~~,~~~c = \left(\frac{2\mu}{m}\right)^{1/2}~~~,~~~
D = \frac{\rho_{\infty}}{\pi}\left(\frac{2m\mu}{\hbar^2}\right)^{1/4}.
\label{Dval}
\end{equation}
Notice also  how the chemical potential entered only via the Fermi
velocity as anticipated (excluding the $\mu$ dependence of the
pre-factor $D$).

We now consider the Fourier transform of (\ref{cardy}) to obtain the
corresponding $G_L (k, i \omega_n)$ at the Matsubara frequencies
along the imaginary frequency axis
\begin{equation}
G_L (k, i\omega_n ) = \int_{-\infty}^{\infty} dx \int_0^{\hbar/(k_B T)}
d \tau e^{-i(kx - \omega_n \tau)} G_L (x, \tau)
\end{equation}
Given the scaling form of $G_L (x,\tau)$  it follows that the two
integrals involved in the transform lead us to the scaling form
\begin{equation}
G_L^R ( k, \omega) = \frac{D}{c} \left(\frac{\hbar c}{k_B
T}\right)^{2-\eta} {\cal A}_L \left( \frac{\hbar \omega}{k_B T},
\frac{\hbar c k}{k_B T} \right),
\label{scalelut}
\end{equation}
where ${\cal A}_L$ is a completely universal scaling function, dependent
only upon the value of $\eta$. We will determine
${\cal A}_L$ in closed form below.

Now if we take the limit $\mu/k_B T \rightarrow \infty$ at {\em fixed\/}
$\omega/k_B T$ and $\hbar c k/k_B T$ (while satisfying (\ref{condmu}) of
course), the system is described simultaneously by the Luttinger liquid
result (\ref{scalelut}) and the zero scale-factor universality of
(\ref{scalegr}). Comparing these two results and using (\ref{Dval}) we
obtain immediately the compatibility condition between
the reduced scaling function ${\cal A}_L$ at $\eta=1/2$ and
the scaling function ${\cal A}$
\begin{equation}
{\cal A}_L ( \overline{\omega} , \overline{k} ) |_{\eta=1/2}
= \frac{\pi}{\sqrt{2} \rho_{\infty}} \lim_{t\rightarrow\infty}
\frac{1}{\sqrt{t}} {\cal A} \left( \overline{\omega},
\frac{\overline{k}}{2\sqrt{t}}, t \right)
\label{acompat}
\end{equation}
We are using here, as before, the notation $\overline{\omega} \equiv
\hbar\omega/(k_B T)$, $\overline{k} \equiv \hbar c k/(k_B T)$
and $t \equiv \mu/(k_B T)$.
We have demanded here that $\mu$ enters ${\cal A}_L$ only through $c$.
Notice that the $t \rightarrow \infty$ limit is taken at fixed
$\overline{\omega}$, and as we have noted before and shall see
explicitly below, the $\overline{\omega} \rightarrow \infty$ limit of
${\cal A}_L$ does not agree with the $\overline{\omega} \rightarrow
\infty$ limit of ${\cal A}$ which was implicit in (\ref{freeparticles})

The remainder of this subsection is devoted to obtaining explicit
results for ${\cal A}_L$ and associated scaling functions
in Minkowski space. There are two ways to proceed.

The first approach is to perform the transform for complex (Matsubara)
frequencies and then  make the substitution $i\omega_n = \omega$ to obtain
$-G^R(\omega)$. The transforms are tedious to perform but the interpretation of
the results is instructive and we shall do so soon.
 The general principles we learn about analytic continuation into the complex
plane are usually illustrated with trivial examples (i.e., noninteracting
propagators) and here we have one of the few  nontrivial and hence instructive
cases.

The second approach is peculiar to this problem and relevant because conformal
invariance methods always give the correlations in coordinate and not momentum
space. Thus one can continue the results from imaginary to real time first and
then take the transform. That calculation may be found in Section 3.3 of
Ref~\cite{spinless} and has its own pedagogical value.

Returning to the first approach, we used the identity
\begin{equation}
X^{-\eta /2} = \frac{1}{\Gamma(\eta /2)}
\int_0^{\infty} d \lambda \lambda^{\eta/2 -1} e^{-\lambda X}
\end{equation}
to put the $\cosh$ and $\sin$ terms in (\ref{cardy}) up in the exponent.
The $x$ and $\tau$ integrals were then analytically performed, followed
finally by the $\lambda$ integration. After using (\ref{scalelut}), the
final result
gave us values of the scaling function ${\cal A}_L$ at the Matsubara
frequencies along the imaginary
frequency axis. We found
\begin{equation}
{\cal A}_L \left( i\overline{\omega} , \overline{k}
\right) = \frac{\pi^{\eta-1}}{2^{2-\eta}} \frac{\Gamma\left({\displaystyle
1-\frac{\eta}{2}}\right)}{
\Gamma\left({\displaystyle \frac{\eta}{2}}\right)}
\frac{ \Gamma \left( {\displaystyle \frac{\eta}{4} + \frac{|\overline{\omega}|
+ i \overline{k}}{
4 \pi }} \right)
\Gamma \left( {\displaystyle \frac{\eta}{4} + \frac{|\overline{\omega}| - i
\overline{k}}{
4 \pi }} \right)}{
\Gamma \left( {\displaystyle 1-\frac{\eta}{4} + \frac{|\overline{\omega}| + i
\overline{k}}{
4 \pi }} \right)
\Gamma \left( {\displaystyle 1-\frac{\eta}{4} + \frac{|\overline{\omega}| - i
\overline{k}}{
4 \pi }} \right)}
\label{resal}
\end{equation}
This result was obtained earlier by Schulz~\cite{schulz} and
Shankar~\cite{spinless} in a different context but not analyzed
in any detail.

First, the above function ${\cal A}_L(|\overline{\omega}|)$ specifies our
knowledge at positive and negative Matsubara points. Our goal is to construct
the physical real frequency correlation function and its singularity structure
from it.  As it stands, ${\cal A}_L$ can only  be used for numerical purposes
and not for studying analytic structure since $ |\overline{\omega}|$ is neither
analytic nor anti-analytic. In other words we can use ${\cal A}_L$ to
calculate values of the putative function at the Matsubara points. The mod
symbol tells us that the function we are seeking has the same value  at any
postive Matsubara point and its negated image.

Now, on the real frequency axis we have a retarded correlation function, which
we assume is well defined. Since the factor $e^{i\omega t}$ converges in the
upper half-plane (UHP) the function on the real axis has an analytic extension
to the UHP which is free of singularities. Its values at the Matsubara points
 $\overline{\omega} = 2 \pi n$, with $n$ an integer are given by ${\cal A}_L$.
Such a function is readily found: simply drop the mod symbol on
$\overline{\omega}$ in the formula for ${\cal A}_L$.
Let us call the function (with the mod symbol dropped) ${\cal
A}_{L}^{UHP}(\overline{\omega})$.This ${\cal A}_{L}^{UHP}(\overline{\omega})$
is the unique  analytic function (with good behaviour in the UHP) determined by
our knowledge at postive Matsubara frequencies.  Being an analytic function it
has a continuation to the lower half-plane (LHP) which is however not
guaranteed
to be free of singularities or to have anything to do with the original
problem.
In particular the poles   that the $\Gamma$ functions have in the
LHP are not germane to the physical response function.
In fact  this continuation to the LHP of  ${\cal
A}_{L}^{UHP}(\overline{\omega})$ does not even agree with the data we have in
Eqn.(\ref{resal}) for negative Matsubara points:  since ${\cal
A}_{L}^{UHP}(\overline{\omega}) \ne {\cal A}_{L}^{UHP}(-\overline{\omega})$ it
is not invariant under the change of sign of frequency as the given data are.
However, there is an analytic function which will duplicate the given data in
the LHP:
it is obtained by replacing $|\overline{\omega}|$ by $-\overline{\omega}$ in
Eqn.(\ref{resal}).  Such a function,
${\cal A}_{L}^{LHP}(\overline{\omega})$, satisifes
\begin{equation}
{\cal A}_{L}^{LHP}(\overline{\omega}) = {\cal
A}_{L}^{UHP}(-\overline{\omega}). \end{equation}
This function will agree with $\cal{A}_L$ of Eqn.(\ref{resal}) at points with
negative Matsubara frequencies and be
 free of singularities in the LHP.  However  its poles in the UHP have no
physical significance.
Thus the function at real frequencies is the limit of two different functions
as we approach the real axis from above or below. {\em The true
singularities of the physical response function are due to the mismatch
of these two functions and not due to the poles they have in regions
where they no longer represent the physical function or a continuation
of it. } So we must consider the difference between these two functions
${\cal A}_{L}^{UHP}(\overline{\omega})$ and ${\cal
A}_{L}^{LHP}(\overline{\omega})$ on the real  frequency axis. It is
readily shown that the two functions are conjugates of each other there
so that the discontinuity is just twice  the imaginary part.  The
physical response function's real singularity is therefore a cut and not
poles.  This is reasonable since a theory with gapless excitations on
top of the ground state (and hence a cut at zero temperature) cannot
lose its spectral weight in these regions by the inclusion of higher
states in the thermal average at finite temperatures. (The converse is
possible: a cut free region on the real axis at $T=0$ can close up at
$T \ne 0$.)

In Fig~\ref{luta} we plot $-\mbox{Im} {\cal A}_{L}
(\overline{\omega} , \overline{k})/\overline{\omega}$
as a function of $\overline{\omega}$
and at a representative set of values of $\overline{k}$.
Notice that for large $\overline{k}$ there is a well defined peak at
$\overline{\omega} \sim \overline{k}$: this is a signature of the propagating
modes in the Luttinger liquid ground state and represents the
behavior of the intermediate region $k_B T \ll \omega \ll \mu$ in
Fig~\ref{omega}. At smaller values of $\overline{k}$ notice that the peak in
Fig~\ref{luta} remains at $\overline{\omega} = 0$. This is the $z=1$ quantum
relaxational behavior (Fig~\ref{omega}) where the strong interaction between
the thermally excited Luttinger modes has left only overdamped excitations.

We now consider a couple of other experimental observables, related to
local and equal-time correlations respectively.

\subsubsection{Local Green's function}

The local Green's function, $G^R_{\ell}$ was defined in (\ref{defloc}).
In the Luttinger liquid regime, we can deduce that,
provided $\omega \ll \mu$, this observable satisfies the
scaling form
\begin{equation}
\mbox{Im} G^R_{\ell L}
(\omega ) = -\mbox{sgn} (\omega ) \frac{D}{c^{\eta}}
|\omega|^{\eta-1} F_{L} \left( \frac{\hbar\omega}{k_B T} \right)
\label{scalelutf}
\end{equation}
where $F_{L}$ is a universal function, specified completely
by the function
${\cal A}_L$ in (\ref{defloc}). Moreover, as the Luttinger-liquid
criticality has a particle-hole symmetry, $F_L$ must be an even,
positive function of $\overline{\omega}$.
As already noted,
we expect on general grounds that $\mbox{Im} G^R_{\ell} ( \omega )
\sim \omega$ for small $\omega$
at finite $T$: therefore
$F_{L} ( \overline{\omega} ) \sim |\overline{\omega}|^{2-\eta}$
at small $\overline{\omega}$.
We also note that for $\eta > 1$
the {\em real\/} part of
the local Green's function will not
satisfy an analogous
because the integral in (\ref{defloc}) is then
dominated by large momentum contributions.

There is again a compatibility
condition between the Luttinger liquid scaling function $F_{L}$ and
the scaling function $F$ in (\ref{deff}) quite analogous to that for
${\cal A}_L$, ${\cal A}$ in (\ref{acompat}); we have
\begin{equation}
F_L ( \overline{\omega} ) |_{\eta=1/2} = \frac{\pi}{\rho_{\infty}}
\lim_{t \rightarrow \infty} F ( \overline{\omega}, t)
\label{fcompat}
\end{equation}
As before, the limits $\overline{\omega} \rightarrow \infty$ and $
t \rightarrow \infty$ do not commute, and the $\overline{\omega}
\rightarrow \infty$ limit of the exact $F_L$ computed below will not
agree with that of $F$ in (\ref{freeparticles})

Let us finally present the exact computation of $F_L$.
We use the result (\ref{cardy}) at $x=0$, Fourier transform to Matsubara
frequencies,
analytically continue and take the imaginary part to obtain the following
result for $F_L$
\begin{equation}
F_L \left( \overline{\omega} \right)
= \left| \overline{\omega} \right|^{1-\eta}
\pi^{\eta-1/2} \sinh \left( \frac{|\overline{\omega}|}{2} \right)
\frac{\left| \Gamma \left( {\displaystyle \frac{\eta}{2} -
\frac{i|\overline{\omega}|}{
2 \pi }}\right)\right|^2}{
\Gamma\left({\displaystyle
\frac{1+\eta}{2}}\right)
\Gamma\left({\displaystyle
\frac{\eta}{2}}\right)}
\label{resfl}
\end{equation}
A plot of this function is shown in Fig~\ref{lutf}.

For small
$\overline{\omega}$ we have
\begin{equation}
F_L ( \overline{\omega} ) =
 \frac{\pi^{\eta-1/2} \Gamma \left( {\displaystyle \frac{\eta}{2}}
 \right)}{2 \Gamma \left( {\displaystyle \frac{1+\eta}{2}} \right)}
 |\overline{\omega}|^{2-\eta} ~~~;~~~|\overline{\omega}| \rightarrow 0.
\end{equation}
This is the behavior characteristic of the $z=1$ quantum-relaxational
regime of Fig~\ref{omega}. It is expected that the
 limits $\overline{\omega} \rightarrow 0$ and $t
\rightarrow \infty$ {\em do commute\/}, so combined with (\ref{fcompat}),
the above result gives us the small $\overline{\omega}$ behavior of
$F(\overline{\omega} , t)$ at large values of $t$.

In
the opposite limit of large $\overline{\omega}$ we crossover to the critical
correlations of the Luttinger liquid ground state in which case
\begin{equation}
F_L ( \overline{\omega} ) = \frac{2^{1-\eta} \pi^{3/2}}{
\Gamma\left({\displaystyle
\frac{1+\eta}{2}}\right)
\Gamma\left({\displaystyle
\frac{\eta}{2}}\right)}
{}~~~;~~~|\overline{\omega}| \rightarrow \infty
\end{equation}
This last result can also be obtained by a Fourier transform
of the relativistic zero temperature correlator.

\subsubsection{Structure Factor}
The two structure factors $S_{+-}(k)$ and $S_{-+}$ of the antiferromagnet
were defined in Eqn (\ref{chipm}). The Luttinger liquid behavior has
particle-hole symmetry so in this regime, the two structure factors are
essentially equal and will be denoted by by the common value $S_L (k)$.
The scaling form for $S_L ( k)$ follows from (\ref{introz}), (\ref{chipm})
and the scaling of $G^R_L$ in (\ref{scalelut}):
\begin{equation}
S_L(k) = Z D \left( \frac{\hbar c}{k_B T} \right)^{1-\eta} {\cal B}_{L}
\left( \frac{\hbar c k}{k_B T} \right),
\label{scalelutb}
\end{equation}
where the constant $Z$ was introduced in (\ref{introz}) and ${\cal B}_{L}$
is a universal function obtained below. There
is a compatibility
condition between the Luttinger liquid scaling
function ${\cal B}_{L}$ and the scaling function ${\cal B}_{+-}$ in
(\ref{defb1})
which is quite analogous to that for ${\cal A}_L$, ${\cal A}$ in
(\ref{acompat}):
\begin{equation}
{\cal B}_L ( \overline{k}) |_{\eta = 1/2} = \frac{\pi}{2 \rho_{\infty}} \lim_{t
\rightarrow \infty} \frac{1}{\sqrt{t}} {\cal B}_{+-} \left(
\frac{\overline{k}}{2\sqrt{t}},  t \right)
\end{equation}
Using the other scaling function ${\cal B}_{-+}$ on the right-hand side would
yield, from (\ref{bb1}), an identical result. Again the limits $t\rightarrow
\infty$ and $\overline{k} \rightarrow \infty$ are not expected to commute.

Finally, the exact determination of ${\cal B}_L$:
we simply perform a spatial Fourier transform of
(\ref{cardy}) at $\tau = 0$; we obtain in this manner
\begin{equation}
{\cal B}_{L} \left ( \overline{k} \right)
= \pi^{\eta-1/2} \frac{\Gamma\left({\displaystyle
\frac{1-\eta}{2}}\right)}{
\Gamma\left({\displaystyle \frac{\eta}{2}}\right)}
\left|
\frac{ \Gamma \left( {\displaystyle \frac{\eta}{2} + \frac{i \overline{k}}{
2 \pi }} \right)}{
\Gamma \left( {\displaystyle \frac{1}{2} + \frac{ i \overline{k}}{
2 \pi }} \right)} \right|^2
\label{resbl}
\end{equation}
This result is illustrated in Fig~\ref{lutb}. For small $\overline{k}$,
${\cal B}_L$ reaches a constant whose value is easily obtainable from
(\ref{resbl}). For large $\overline{k}$ we find from (\ref{resbl})
\begin{equation}
{\cal B}_L ( \overline{k} ) = \frac{2^{1-\eta} \pi^{1/2}
\Gamma\left({\displaystyle
\frac{1-\eta}{2}}\right)}{
\Gamma\left({\displaystyle
\frac{\eta}{2}}\right)}
\frac{1}{\overline{k}^{1-\eta}}
{}~~~;~~~|\overline{k}| \rightarrow \infty
\end{equation}
Again this last result could have also been obtained by direct computation at
$T=0$.

\section{CONCLUSIONS}
\label{conc}
This paper has studied the universal, finite temperature properties of a
dilute Bose gas with repulsive interactions in dimensions less than or equal
to 2. In the vicinity of the $T=0$ onset at zero chemical potential, $\mu$, it
was argued that the leading scaling properties obey, for $d < 2$, a hypothesis
of zero scale-factor universality. This means that the entire two-point
correlator is a universal function of just $\mu$, $T$ and the bare boson mass
$m$.

The main motivation behind this study is the mapping onto it of the
properties of quantum-disordered antiferromagnets in a finite field.
In particular, in $d=1$, Haldane gap antiferromagnets undergo a magnetization
onset at a critical field which is expected to be in the universality class
of the Bose gas transition. Applicability of our theory requires that
there be no spin anisotropy in the plane perpendicular to the applied field.
Most materials do have some anisotropy - in this case we would require that
the temperature, $T$, be larger than the anisotropy gap, before applying our
results.

We have described the rather complicated properties of numerous scaling
functions, which may rather difficult to disentangle experimentally.
A useful starting point for neutron scattering experiments appears to be the
following. Perform the experiment somewhere in the Luttinger liquid region
where the absolute value of the scattering cross-section is also the largest.
Measure the local susceptibility $G^R_{\ell} (\omega )$ and see if collapses
onto the scaling form (\ref{deff}). For large $\mu /k_B T$, we have a
rather complete picture of the scaling function $F$: for $\omega$ smaller than
or around $k_BT$, we can deduce $F$ from (\ref{fcompat}) and (\ref{resfl}),
while for extremely large $\omega$ we can use (\ref{freeparticles}).

Another possible application of our results may be to quantum-disordered
antiferromagnets in $d=2$. By measuring the ground-state magnetization in
a field, and comparing the result to (\ref{muvalh}), (\ref{mvaln}) and
(\ref{popovres}) it may be possible to determine the spin $S_z$ of the
elementary excitations above the ground state. Of course, we would also need
an independent determination of the quasi-particle mass $m$.

\acknowledgments

We thank I. Affleck, I. Gruzberg,
B.I. Halperin, P. Hohenberg, and V. Korepin for helpful discussions
and correspondence.
The research was supported by NSF Grants No. DMR-9120525 and
DMR-9224290.

\appendix

\section{Magnon operators}
\label{shankar}

It is clear that the predominant coupling of neutrons will be to
the antiferromagnetic order parameter $\phi_{\alpha}$, with $\alpha=x,y,z$.
In a zero-field spin-fluid phase with confined spinons, this order parameter
corresponds to a real, massive, bosonic triplet.
In this appendix, we want to explore in some more detail the relationship
between $\phi_{\alpha}$ and the complex bosonic field, $\Psi$ in
in Eqn.(\ref{coherent}). We note that the field theory (\ref{coherent})
will also describe the magnetization onset transition in antiferromagnets
with deconfined spinons; however, in this case, the relationship between the
neutron scattering cross-section and the field $\Psi$ will be quite
different~\cite{triangle}, and will not be considered in this paper
explicitly.

For simplicity, we consider $d=1$, although the analysis is quite general.
First we expand the real triplet, $\phi_{\alpha}$,   in terms of magnon
creation
and destruction operators as usual:
\begin{equation}
{\phi}_{\alpha} (x) = \int_{-\infty}^{\infty} {dk \over 2\pi}{1 \over
\sqrt{2\omega_k}} [ a_{\alpha} (k) e^{ikx} + a^{\dag}_{\alpha} e^{-ikx}]
\end{equation}
where
\begin{equation}
\omega_k = \sqrt{\Delta^2 + k^2},
\end{equation}
$\Delta$ is the Haldane gap, and $a_{\alpha}$, $a^{\dag}_{\alpha}$ the magnon
destruction and creation operators for each of three polarizations. Let us
focus
on the two combinations
\begin{equation}
{\phi}_{\pm}(x)= {\phi_x \pm \phi_y \over \sqrt{2}} = \int_{-\infty}^{\infty}
{dk \over    2\pi}{1 \over \sqrt{4\omega_k}} [ (a_x (k)\pm i a_y(k)) e^{ikx} +
(a^{\dag}_{x}(k) \pm i a^{\dag}_y (k)) e^{-ikx}] .
\end{equation}
Observe that $\phi_{\pm}$ are adjoints of each other, but commute with each
other. Let us now consider an effective theory for energies far below the
Haldane gap. In this case we can make the replacement
\begin{equation}
\omega_k \simeq \Delta
\end{equation}
and obtain
\begin{eqnarray}
\phi_{\pm}(x)  &=& \int_{-\infty}^{\infty} {dk \over 2\pi}{1 \over
\sqrt{4\Delta}} [ (a_x (k)\pm i a_y(k)) e^{ikx} + (a^{\dag}_{x}(k) \mp i
a^{\dag}_y (k)) e^{-ikx}] \\
 &\equiv& {1 \over \sqrt{2\Delta}} [ \Psi_{\pm }(x) + \Psi^{\dag}_{\pm}(x) ]
\end{eqnarray}
where $\Psi_{\pm }(x)$ destroys a spin $\pm 1$ magnon at $x$ while
$\Psi^{\dag}_{\mp}(x) $ creates a spin $\mp 1$ magnon at $x$.
Suppose we next argue that, when the applied uniform field is near its critical
value,
only the spin up magnon (very light) will either be easily created or
destroyed,
so that we may drop the spin down creation and destruction  operator in the
above expressions.  Then we obtain
\begin{eqnarray}
\phi_{+}&=& {1 \over \sqrt{2\Delta} } \Psi_{+ }(x) \\
\phi_{-} &=& {1 \over \sqrt{2\Delta}}\Psi^{\dag}_{+ }(x) .
\end{eqnarray}
Observe that now (up to a scale factor) $\phi_{\pm}$, which were previously
commuting, {\em are now canonically conjugate fields.} This is just like in the
Hall effect wherein $x$ and $y$ , which are commuting coordinates in the full
Hilbert space become conjugates in the lowest Landau level. It is also clear
from the discussion that the field $\Psi$ in the coherent space integral in
Eqn.(\ref{coherent}) is precisely this complex conjugate pair.

\section{Boson Hubbard model in one dimension}
\label{canonical}
Consider bosons $b_i$ moving on the sites, $i$ of a chain described
by the Hamiltonian
\begin{equation}
H = -w \sum_{i} \left( b_i^{\dagger} b_{i+1} + b_{i+1}^{\dagger} b_i
-2 b_i^{\dagger} b_i \right) + \sum_i \left( V n_i ( n_i -1) -
\mu n_i \right)
\end{equation}
where $n_i = b_i^{\dagger} b_i$ is the number operator, $w$ is the
hopping matrix element, and $V$ is the on-site repulsion between the
bosons. In the limit of large $V$, states with more than one boson
on a site will only occur rarely, and it should pay to restrict the
Hilbert space by projecting out such states. However, the elimination will
generate  a residual interaction of order $w^2 /V$ between
the states on the restricted space. This interaction can be determined by the
usual second-order perturbation theory and leads to the effective
Hamiltonian
\begin{eqnarray}
H_e &=& -w \sum_{i} \left( b_i^{\dagger} b_{i+1} + b_{i+1}^{\dagger} b_i
-2 b_i^{\dagger} b_i \right) -\mu \sum_i
n_i \nonumber \\
&~&~~~~~~~~~~~-\frac{2w^2}{V} \sum_i \left( 2 b_i^{\dagger}
b_{i+1}^{\dagger} b_{i+1} b_i + b_i^{\dagger} b_{i-1}^{\dagger}
b_{i+1} b_i + b_i^{\dagger} b_{i+1}^{\dagger} b_{i-1} b_i \right)
\label{heffec} \end{eqnarray}
We reiterate that $H_e$ is non-zero only on states with at most one
boson per site.
Notice now that this reduced Hilbert space is identical to that of
spinless fermions. The transformation between the $b_i$ and the
spinless fermion operators $f_i$ is of course the Jordan-Wigner
mapping
\begin{equation}
b_i = \prod_{j<i} ( 1- 2 f_j^{\dagger} f_j ) f_i
\label{jw}
\end{equation}
We now insert (\ref{jw}) in (\ref{heffec}) and take the continuum
limit with $f_i = \sqrt{a} \Psi_F (x=ia)$, $w = \hbar^2 /(2 m a^2)$
($a$ is the lattice spacing) and obtain
\begin{equation}
 H_F = \int dx \left[ \Psi_F^{\dagger} \left( - \frac{\hbar^2}{2m}
 \frac{d^2}{d x^2} - \mu \right) \Psi_F - \frac{8w^2 a^3}{V} \frac{d
\Psi_F^{\dagger}}{dx} \Psi_F^{\dagger} \Psi_F \frac{d \Psi_F}{dx}
\right]
\end{equation}
It is now clear by power counting that the four-Fermi coupling term is
clearly an irrelevant perturbation to the $\mu=T=0$ fixed point.
It is in fact also not difficult to see that all interactions between
the fermions are irrelevant, and that this result is not special to the
Boson Hubbard model considered here. The key point is of course that a
term like $\Psi_F^{\dagger} \Psi_F^\dagger \Psi_F \Psi_F$, which is the
only interaction term which is
relevant by power counting about the free fermion fixed point at
$\mu=0$, vanishes identically because of the fermion anticommutation
relations.

The significance of the four-Fermi coupling changes when we
consider the scaling dimensions of operators about the Luttinger liquid
fixed points. In this case we decompose the fermion field into left-
($\Psi_L$) and right- ($\Psi_R$) moving excitations with a linear
dispersion, and obtain the long-wavelength Hamiltonian
\begin{equation}
H_L = \int dx \left[ \hbar c \left( \Psi_R^{\dagger} \frac{d\Psi_R}{dx}
- \Psi_L^{\dagger} \frac{d\Psi_L}{dx} \right) - \frac{32w^2 a^3 k_F^2}{
V} \Psi_{R}^{\dagger} \Psi_L^{\dagger} \Psi_L \Psi_R \right]
\end{equation}
where $c = \hbar k_F /m$ and the Fermi wavevector $k_F$ is given
by $\hbar^2 k_F^2 /(2m) = \mu$. Performing power-couting on the
$z=1$ free field part of $H_L$ we now find that the four-Fermi coupling
is now marginal. Note however that the coefficient of this four-Fermi
coupling is suppressed by a factor of $k_F^2$, which vanishes as
one approaches the $z=2$ critical end-point. By the
usual logarithmic perturbation theory at $T=0$ we can determine that
the four-fermi interaction
modifies the exponent $\eta$ of Section~\ref{lutliq} by
\begin{equation}
\eta = \frac{1}{2} -  \gamma \frac{\sqrt{w\mu}}{V},
\end{equation}
where $\gamma$ is a numerical constant of order unity.
Notice, as expected, that the correction to $\eta$ vanishes as
$\mu \rightarrow 0$. It is also apparent that the impenetrable limit
$V \rightarrow \infty$ is equivalent to the vanishing density ($\mu
\rightarrow 0$) limit.

At finite $T$, there will be corrections to the correlators with terms
like $(\sqrt{t\mu}/V) \log(\mu/k_B T)$. These cannot be neglected when
\begin{equation}
k_B T < \mu \exp\left(- \gamma^{\prime}\frac{V}{\sqrt{w\mu}}\right)
\label{violation}
\end{equation}
where $\gamma^{\prime}$ is of order unity. Zero scale-factor
universality is thus violated for arbitrarily small $\mu$, when $T$ is
smaller still and satisfies (\ref{violation}). Notice however that the
boundary specified by (\ref{violation}) lies well below $k_B T \sim \mu$
crossover to the Luttinger liquid regime (Fig~\ref{figd1}).

\section{Impenetrable Bose Gas in one dimension}
\label{itsetal}
Its {\em et. al.\/}~\cite{korep1} have recently obtained some exact
asymptotic results for
the equal-time boson Green's function of the $d=1$ impenetrable Bose gas.
Recall that this model is precisely the scaling limit describing the $z=2$
quantum phase transition with zero scale-factor universality. In this Appendix,
we show how the requirement that the scaling functions be analytic in
$\mu /k_B T$, can lead to a considerable simplification of their results.

The asymptotic results of Its {\em et.al.\/} can be written in the form
\begin{equation}
G(x, \tau=0^{-1}) = \frac{\sqrt{2mk_B T}}{\hbar} A\left( \frac{\mu}{k_B T}
\right) \exp \left[ -  \frac{\sqrt{2mk_B T}}{\hbar}
f\left(\frac{\mu}{k_B T}\right) x\right]~~~~\mbox{as $x \rightarrow \infty$}.
\label{gkorrep}
\end{equation}
where $A(t)$ and $f(t)$ are functions to be determined (as before $t \equiv
\mu/k_B T$).
{}From the arguments in Section~\ref{onedimension},
it is clear that $A(t)$ and $f(t)$ are
also universal crossover functions of the $\mu=0$, $T=0$, quantum
phase transition the repulsive, $d=1$ Bose gas with {\em arbitrary\/},
short-range interactions.
Its {\em et. al.\/} obtained two separate, closed-form, integral expressions
for $f_{\pm}(t)$ and $A_{\pm} (t)$ valid respectively for $t > 0$ and $t < 0$.
The two expressions were quite distinct and there appeared to be no
straightforward relationship between them.

Here, we point out that the absence of any singularity in the finite $T$ Bose
gas, in fact requires that $f(t)$  and $A(t)$ be analytic for all finite, real
values of $t$.  In other words, the functions $f_{+} (t)$ and $f_{-} (t)$
must be analytic continuations of each other (similarly for $A_{+} (t)$
and $A_{-} (t)$). We have in fact succeeded in proving that the
expression of Its {\em et. al.\/} for $f_+ (t)$ is in fact the analytic
continuation of their result for $f_- (t)$. We have been unable to establish a
similar result for $A_{\pm} (t)$, but have performed numerical tests
on their expressions, which leave essentially no doubt that
$A$ is also analytic.

\begin{figure}
\caption{The scaling functions $f(t)$ and $A(t)$ ($t = \mu / (k_B T)$)
of the $d=1$ Bose gas
defined by Eqns (\protect\ref{gkorrep}) and (\protect\ref{afint})}
\label{korepfig}
\end{figure}
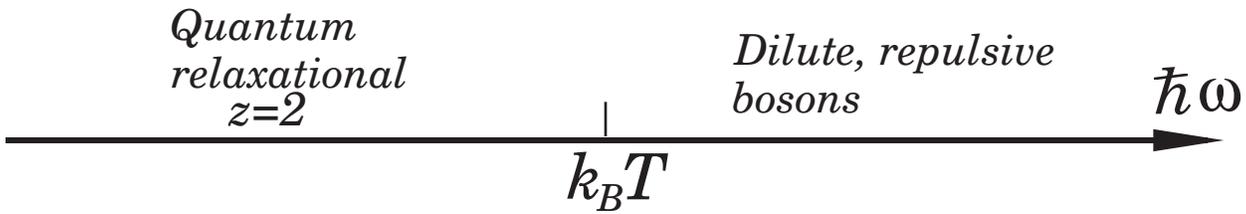
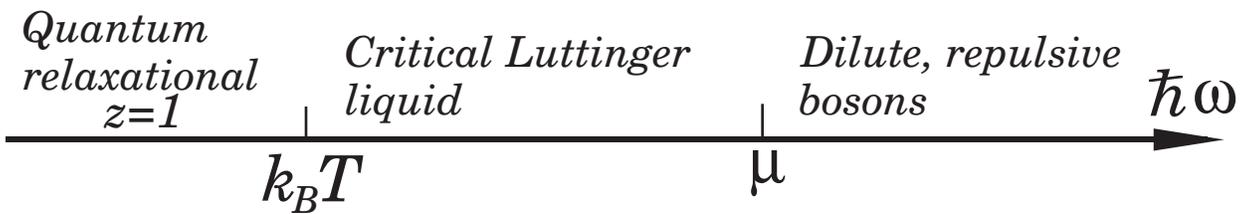
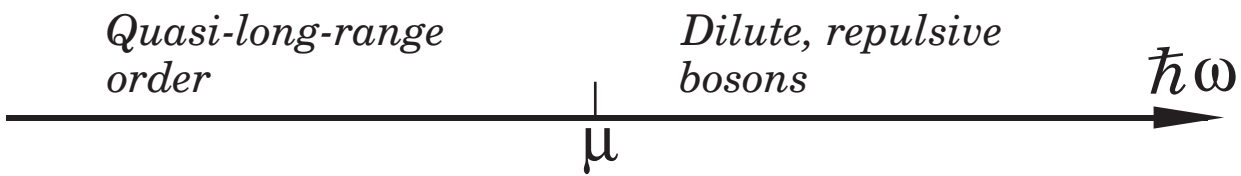

With the help of the above considerations, it is possible to deduce from
Ref~\cite{korep1} a simple closed-form result for $f(t)$ and $A(t)$ which is
valid for {\em all\/} $t$
\begin{eqnarray}
f(t) &=& 1 + \frac{1}{\pi}
\int_0^{\infty} dy \log \left( \frac{(e^{y^2 - t} + 1)(y^2 - t)}{(e^{y^2 - t} -
1)(y^2 + 1)} \right) \nonumber \\
A(t) &=& \frac{\rho_{\infty}}{\sqrt{\pi}}
\exp\left( - 2
\int_t^{\infty} d y \left( \frac{df(y)}{dy} \right)^2 \right).
\label{afint}
\end{eqnarray}
The value of the constant $\rho_{\infty}$ was obtained by
matching to the $T=0$ result of Vaidya and Tracy~\cite{vaidya}
\begin{equation}
\rho_{\infty} =
\pi e^{1/2} 2^{-1/3} A_G^{-6} = 0.924182203782\ldots,
\end{equation}
$A_G$ being Glaisher's
constant~\cite{ttwu}. Note that the analyticity of $f$ and $A$ for all real $t$
is manifest. We have plotted the functions $f(t)$ and $A(t)$ in
Fig~\ref{korepfig}. They obey the
asymptotic limits
\begin{eqnarray}
f(t) &=& \left\{ \begin{array}{cc}
{\pi}/({4\sqrt{t}}) &~~t \rightarrow\infty \\
\sqrt{-t} &~~t \rightarrow-\infty
\end{array} \right. \nonumber\\
A(t) &=& \left\{ \begin{array}{cc}
{\rho_{\infty}}/{\sqrt{\pi}} &~~t \rightarrow\infty \\
{1}/({2\sqrt{-t}}) &~~t \rightarrow-\infty
\end{array} \right. ,
\end{eqnarray}
Both asymptotic limits of $f(t)$, and the $t \rightarrow +\infty$ limit of
$A(t)$, can be obtained directly from (\ref{afint}). The $t\rightarrow -\infty$
limit of $A(t)$ is more difficult to obtain from (\ref{afint}), and we used
instead the second expression for $A(t)$ in Ref~\cite{korep1}. Demanding that
these two methods of obtaining the limit be identical in fact provides one with
an independent derivation of the value of the constant $\rho_{\infty}$ !

We also recall~\cite{korep1} that
\begin{equation}
f(0) = \frac{\zeta(3/2)}{\sqrt{\pi}} \left( 1 - \frac{1}{2 \sqrt{2}} \right)
= 0.95278147061075\ldots
\end{equation}

We have already noted that taking
a Fourier transform of the asymptotic results
(\ref{gkorrep}) to obtain the structure
factor, yields results which compare very poorly with numerically exact
results of Fig~\ref{figb1}.

Finally, we note that the requirement of analyticity as a function of $t$
should apply to essentially all of the equal-time and unequal-time results
of Its~{\em et. al.\/}~\cite{korep1,korep2}. In every case, they have obtained
separate expressions for $t <0$ and $t > 0$: proving that these are analytic
continuations of each other will lead to highly non-trivial checks on the
results, and should also produce some fascinating mathematical identities.

\begin{figure}
\caption{Phase diagram in $d=1$. The dashed lines indicate crossovers.}
\label{figd1}
\end{figure}
\begin{figure}
\caption{Phase diagram in $d=2$. The dashed line is a crossover while the
full line is a Kosterlitz-Thouless phase transition. The field $H$ has
absorbed a factor of $g \mu_B$. The location of the Kosterlitz-Thouless
transition is determined from Refs~\protect\cite{popov,fisher}}
\label{figd2}
\end{figure}
\begin{figure} \caption{Properties of the different regimes of
Fig~\protect\ref{figd1} and~\protect\ref{figd2} as a function of the
measurement frequency $\omega$ (the wavevector $k\approx 0$). All crossovers
are described by the universal scaling function ${\cal A}$. The crossover for
$\mu \gg k_B T$ in $d=1$ near $\hbar \omega \sim k_B T$ is also described by
the Luttinger liquid scaling function ${\cal A}_L$}
\label{omega}
\end{figure}
\begin{figure}
\caption{Exact results
for the scaling function ${\cal B}_{-+} (r,t)$ for the structure factor
(Eqn (\protect\ref{defb1})) in $d=1$. We have $r = \hbar k/\protect\sqrt{2 m
k_B T}$,
and $t=\mu / (k_B T)$. We have chosen some representative
some representative values of $t$; a computer program to evaluate ${\cal
B}_{-+}$ for arbitrary $t$ is available from the authors}
\label{figb1}
\end{figure}
\begin{figure}
\caption{Exact values of the scaling function $- \mbox{Im}
{\cal A}_L (
\overline{\omega}, \overline{k})/\overline{\omega}$ (given in Eqn
(\protect\ref{resal})) for the Green's function in $d=1$
in the Luttinger liquid regime
at $\eta=1/2$ - Eqn (\protect\ref{scalelut}). We have $\overline{\omega} =
\hbar \omega / (k_B T)$,
$\overline{k} = \hbar c k / (k_B T)$. The values for $\overline{k}$ are
3 times larger than those on the graph. Notice how the spectrum evolves from
an overdamped, relaxational peak at small $\overline{k}$ (as for
$\overline{k}=1$) to a damped, propagating mode at
large $\overline{k}$ (as for $\overline{k}=3,5$).}
\label{luta}
\end{figure}
\begin{figure}
\caption{Exact scaling function ${F}_{L} ( \overline{\omega})$
(given in Eqn (\protect\ref{resfl}))
for the imaginary part of the local susceptibility in $d=1$ in the
Luttinger liquid regime at $\eta=1/2$ -
Eqn (\protect\ref{scalelutf}). We have
$\overline{\omega} = \hbar \omega / (k_B T)$.}
\label{lutf}
\end{figure}
\begin{figure}
\caption{Exact scaling function ${\cal B}_{L} ( \overline{k})$
(given in Eqn (\protect\ref{resbl}))
for the structure factor in $d=1$ in the Luttinger liquid regime at $\eta=1/2$
- Eqn (\protect\ref{scalelutb}). We have
$\overline{k} = \hbar c k / (k_B T)$.}
\label{lutb}
\end{figure}
\end{document}